\documentclass[twocolumn,showpacs,preprintnumbers,amsmath,amssymb]{revtex4}

\usepackage{graphicx} 

\begin{document}

\title{Stochastic resolution of the LHC inverse problem}

\author{Csaba Bal\'azs} \email{Csaba.Balazs@sci.monash.edu.au}
\affiliation{School of Physics, Monash University, Melbourne Victoria 3800, Australia}

\author{Dilani Kahawala} \email{kahawala@physics.harvard.edu}
\affiliation{Jefferson Physical Laboratory, Harvard University, Cambridge Massachusetts 02138, USA \vspace{2 mm}}

\date{March 31, 2009}

\begin{abstract}

%

In this work the LHC inverse problem is quantified in the Bayesian context by clarifying the relation between the mapping from the theory parameter space to experimental signature space and the inverse map.  We demonstrate that, after complementing the LHC data by existing astrophysical, collider, and low energy measurements, a simple likelihood analysis is able to significantly reduce the inverse problem.
The presented approach offers a robust, economical, and extendable way to extract theoretical parameters from the LHC, and other experimental, data.

\end{abstract}

 

\maketitle

\section{Introduction \label{sec:Introduction}}


The CERN Large Hadron Collider (LHC) will soon start colliding protons at 14 TeV center of mass energy.  The LHC data is expected to shed light on the origin of mass of elementary particles, and discover precursors of a wider theory beyond the standard particle model (SM).  Whatever this new theory might be, it is expected to be described by a Lagrangian having several unknown parameters.  Our goal is the extraction of these theoretical parameters from experimental data.


The feasibility of this goal was studied quantitatively within the context of the minimal supersymmetric standard model (MSSM) in Ref. \cite{ArkaniHamed:2005px}.  The authors found that a significant proportion of LHC signatures map to the theoretical parameter space with a large uncertainty.  They isolated pairs of points located relatively far from each other in the MSSM parameter space producing indistinguishable LHC signals.  This implies that the LHC would not be able to resolve certain model parameter regions with sufficient precision.  The difficulty was coined as the LHC inverse problem.


Meanwhile, the task of parameter extraction has been attacked from another direction.  Numerous authors perform global fits of supersymmetric models using presently available experimental data \cite{Baer:2003yh, Ellis:2003si, Ellis:2004tc, Ellis:2006ix, Buchmueller:2008qe, Allanach:2005kz, Allanach:2006jc, Allanach:2007qk, Feroz:2008wr, deAustri:2006pe, Roszkowski:2006mi, Roszkowski:2007fd, Roszkowski:2007va, Trotta:2008bp}.  These global analyses typically evaluate the likelihood of certain experimental outcomes assuming the constrained version of the MSSM as the theoretical model, spanned by a few parameters.  The calculated likelihoods determine the probability distributions of each of the theoretical parameters through marginalization.  Even though this procedure maps the signature space to the theoretical parameter space, so far there has been no explicit connection established with the inverse problem.


In this work, we point out that the inverse problem can be rigorously quantified, and simply connected to the likelihood analyses within the framework of Bayesian probability theory.  First, we quantify a map from signature to theory space by introducing the probability that the theoretical parameters are restricted to certain ranges by a given set of experimental data.  Then, we represent the map from theory to signature space by the likelihood that a certain set of data is predicted by a theory with a given set of parameters.  Since Bayes' theorem explicitly connects the two probabilities, the stochastic framework offers us a robust way to attack the inverse problem. 
Utilizing this statistical formulation, we demonstrate a simple way in which the inverse problem can be substantially reduced by using results from the LHC and other colliders, electroweak precision experiments, rare b-decays, and various astrophysical observations.

\section{The inverse problem in the stochastic framework \label{sec:Stochastic}}

In this section, we recapture the problem of theoretical parameter extraction within the context of probability theory.  
We begin by defining a mapping from the experimental signature space to the theoretical parameter space.  In the context of the inverse problem this 
is referred to as the inverse map.

%
Given a theoretical model $T$ and a set of experimental data $D$, we introduce the probability $\mathcal{P}(P|T;D)$ that the theoretical model describes the data with the theory parameters set to values $P$.  
In the theory of probabilities $\mathcal{P}(P|T;D)$ is known as the conditional probability of the occurrence of $P$ provided that $T$ is assumed while $D$ holds.
According to Bayes' theorem, this probability can be simply calculated as
\begin{eqnarray}
\mathcal{P}(P|T;D) = \mathcal{P}(D|T;P) \frac{\mathcal{P}(T;P)}{\mathcal{P}(D)} .
\end{eqnarray}
Here $\mathcal{P}(D|T;P)$ is the likelihood that a certain set of data is predicted by a theoretical model with a specified set of theory parameters.  The probability $\mathcal{P}(T;P)$ gives the a-priori distribution of the parameters within the theory, fixed by only theoretical considerations independently from the data.   Finally, the evidence $\mathcal{P}(D)$ gives the integrated likelihood of the theory $T$ in terms of the data alone.  From unitarity it follows that
\begin{eqnarray}
\mathcal{P}(D) = \int \mathcal{P}(D|T;P) \mathcal{P}(T;P) dP ,
\end{eqnarray}
where the integral extends over the whole parameter space, $P={p_1,...,p_n}$, of the theory.

The likelihood function $\mathcal{P}(D|T;P)$ maps forward, from the theory space to the signature space.  If the data under consideration are independent (as in our case) $\mathcal{P}(D|T;P)$ factorises as:
\begin{eqnarray}
\label{eq:likelihoodProduct}
\mathcal{P}(D|T;P) = \prod_i {\cal L}_i(D|T;P) ,
\end{eqnarray}
where the product is over $i$ different experiments with ${\cal L}_i$ corresponding to the the $i^{th}$ likelihood function.  Each term in the product is formed as a convolution
\begin{eqnarray}
{\cal L}_i(D|T;P) = {\cal L}_{i,e}(D) \otimes {\cal L}_{i,t}(T;P) .
\end{eqnarray}
If the experimental data and theoretical predictions are normally distributed then ${\cal L}_{i,e}$ and ${\cal L}_{i,t}$ are Gaussians and
\begin{eqnarray}
{\cal L}_i(D|T;P) = \frac{1}{\sqrt{2 \pi} \sigma_i} \exp(\chi_i^2(D,T,P)/2) .
\end{eqnarray}
The exponents of the likelihood functions
\begin{eqnarray}
\label{eq:chi2}
\chi_i^2(D,T,P)/2 = (d_i - t_i(P))^2/2 \sigma_i^2 
\end{eqnarray}
are defined in terms of the experimental data $D = \{d_i \pm \sigma_{i,e}\}$ and theoretical predictions for these measurables $\{t_i \pm \sigma_{i,t}\}$.  Independent experimental and theoretical uncertainties combine into $\sigma_i^2 = \sigma_{i,e}^2 + \sigma_{i,t}^2$.

In cases when the experimental data only specify a lower (or upper) limit, the likelihood function can be written in terms of the error function
\begin{eqnarray}
\mathcal{P}_i(D|T;P) = \frac{1}{2} {\rm erf} (\sqrt{\chi_i^2(D,T,P)/2}) .
\end{eqnarray}


In the stochastic context the task of theoretical parameter extraction, the inverse mapping from signature space to parameter space, boils down to the determination of the posterior probability distribution $\mathcal{P}(P|T;D)$ describing the probability that the parameters of a given theory have certain values in light of a certain set of data.  This can be calculated by evaluating the right hand side of Bayes' theorem.  A parameter is well determined by the data if the variance of the $\mathcal{P}(P|T;D)$ distribution is small.  This means that, the inverse problem can be quantified in terms of the likelihood function, the theoretical prior and the experimental evidence.  


To quantify the inverse process we introduce a quantity called the experimental differentiating power, which ties the inverse map to the forward map via Bayes' theorem.  Given two points $P_1$ and $P_2$ in the parameter space of a theory, we define the experimental differentiating power in terms of the ratio of the prior probabilities:
\begin{eqnarray}
{\cal D} = 1 - \mathcal{P}(P_1|T;D)/\mathcal{P}(P_2|T;D) ,
\end{eqnarray}
where we can always arrange that $0 \leq \mathcal{P}(P_1|T;D) \leq \mathcal{P}(P_2|T;D) \leq 1$ ensuring $0 \leq {\cal D} \leq 1$.
Regardless of the form of the theoretical prior $\mathcal{P}(T;D)$ and the experimental evidence $\mathcal{P}(D)$, the above discriminator can be expressed, via Bayes' theorem, as
\begin{eqnarray}
\label{eq:calD}
{\cal D} =  1 - \mathcal{P}(D|T;P_1)/\mathcal{P}(D|T;P_2) .
\end{eqnarray}


According to Eq.(\ref{eq:calD}), if the theory at parameter point $P_1$ fits the data just as well as at point $P_2$, then the data has little differentiating power between the two model points.  Conversely, $\cal D$ can only be large if the data significantly prefer $P_2$ over $P_1$, that is if $\mathcal{P}(D|T;P_2) \gg \mathcal{P}(D|T;P_1)$.  As formulated in Eq.(\ref{eq:calD}), $\cal D$ gives the combined differentiating power of all experiments in the product of Eq.(\ref{eq:likelihoodProduct}).  We could also calculate the differentiating power of a single experiment ${\cal D}_i$ by including only the $i^{th}$ term in evaluating $\mathcal{P}(D|T;P)$ in Eq.(\ref{eq:likelihoodProduct}).  In the next section, we show that $\cal D$ is, indeed, a robust discriminator in the case of the MSSM and it can be used to significantly reduce the degeneracy in the inverse map.

\begin{table*}[tbh]
\caption{
\label{tab:Data}
 Observables used to reduce the inverse problem in the MSSM.  Experimental results are listed under column $d_i \pm \sigma_{i,e}$, and typical uncertainties related to the MSSM calculations are under $\sigma_{i,t}$.}
\begin{ruledtabular}
\begin{tabular}{lrrrr}
 observable & & limit type & $d_i \pm \sigma_{i,e}$ & $\sigma_{i,t}$ \\
\hline
 lightest Higgs mass                               &                   $m_h$ &   lower limit &                  91.0 - 114.4 GeV   \, \cite{Schael:2006cr} &    3.0 GeV        \cite{Frank:2006yh} \\
 lighter chargino mass                             &       $m_{\tilde{W}_1}$ &   lower limit &                         103.5 GeV \, \cite{Abbiendi:2003sc} &    1.0 GeV       \cite{Paige:2003mg} \\
 electroweak precision parameter                   &         $\Delta \rho_0$ & central value & $(4.0^{+8.0}_{-4.0})\times 10^{-4}$    \cite{Amsler:2008zz} & negligible                           \\
 muon anomalous magnetic moment                    &        $\Delta a_{\mu}$ & central value & (29.0 $\pm$ 9.0)$\times 10^{-10}$ \cite{Jegerlehner:2009ry} & negligible                           \\
 rare b-decay branching fraction                   &     $B(b \to s \gamma)$ & central value & (3.50 $\pm$ 0.17)$\times 10^{-4}$    \cite{Barberio:2008fa} &      10 \%      \cite{Misiak:2006zs} \\
 rare b-decay branching fraction                   & $B(B_s \to \mu^+\mu^-)$ &   upper limit &              4.7 $\times 10^{-8}$      \cite{Artuso:2009jw} &      10 \%       \cite{Buras:2002vd} \\
 average abundance of dark matter                  &            $\Omega h^2$ &   upper limit &                            0.1143     \cite{Komatsu:2008hk} &      10 \%    \cite{Belanger:2006is} \\
 spin independent WIMP-proton elastic recoil       &           $\sigma_{SI}$ &   upper limit &                         CDMS 2008       \cite{Ahmed:2008eu} &      20 \%       \cite{Ellis:2008hf} \\
 gamma ray emission flux from the galactic center  &      $\Phi_\gamma^{GC}$ &   upper limit &             1.0 $\times 10^{-10}$    \cite{Morselli:2002nw} &     100 \% \cite{Altunkaynak:2008ry} \\
\end{tabular}
\end{ruledtabular}
\end{table*}

\section{The inverse problem in the MSSM \label{Likelihood}}

In this section, we illustrate the discrimination power of the Bayesian approach using the MSSM.  The authors of Ref. \cite{ArkaniHamed:2005px} found that out of 43026 randomly selected model points 283 pairs had indistinguishable LHC signatures.  Since they were using discrete experimental data (such as number of jets, etc.), the $\cal D$ values that they could have calculated are only 0 or 1.  Their indistinguishable pairs represent models where the differentiating power of the LHC data vanishes.  


Of these 283 pairs, 162 were analyzed by the authors of \cite{Berger:2007yu, Berger:2007ut} to determine the capability of the International Linear Collider (ILC) to resolve the inverse problem.  Assuming a center of mass energy of 500 GeV, only one third (57) of the pairs were found to be distinguishable at 5 $\sigma$ confidence level.


We start with the 283 degenerate MSSM model pairs of Ref. \cite{ArkaniHamed:2005px} and observe that they consist of 384 unique model points.  This implies that some of the models had more than one degenerate partner, indicating that there were larger degenerate groups of models.  In the population of 384 models we determine the degeneracy groups, such that one model only belongs to just one group, every model within a group is degenerate with every other model within the group, and no model in any group is degenerate with a model in a different group.  We then find all possible pairings of models within a particular degeneracy group.  This allows us to increase our population of model pairs from 283 to 802 degenerate pairs.


For our 802 model pairs, we calculate the discriminator $\cal D$ utilizing experimental data beyond the LHC.  We select measurables from a wide range of experiments: collider limits on particle masses, precision electroweak observables, rare decay branching fractions, and astrophysical observations.  All this experimental data is available now as indicated in TABLE \ref{tab:Data}.  

The theoretical model predictions, as the function of the MSSM soft parameters $t_i(P)$, were calculated using a modified version of ISATools and micrOMEGAs \cite{Belanger:2006is} software packages.  (In this work, we only show results from micrOMEGAs.)  In both cases ISAJET 7.71 \cite{Paige:2003mg} was used to calculate the physical mass spectrum without two-loop corrections to exactly match the spectrum of Ref.s \cite{ArkaniHamed:2005px, Berger:2007yu, Berger:2007ut}.  

\begin{figure}[t]
\includegraphics[width=0.47\textwidth,height=0.47\textwidth]{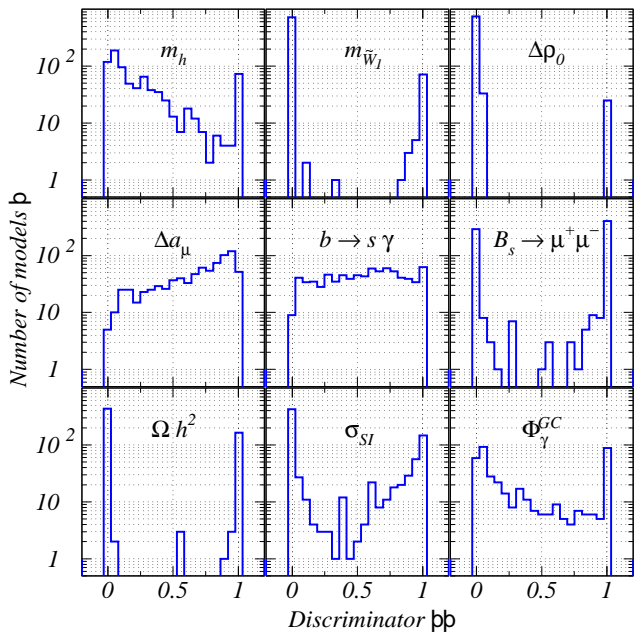}
\caption{\label{fig:D}
Discrimination power of various observables for the case of the MSSM.  The number of model pairs plotted against the discriminator ${\cal D} =  1 - \mathcal{P}(D|T;P_1)/\mathcal{P}(D|T;P_2)$.  Only one observable (a single term) is used in the likelihood functions ${\cal P}$ in each frame.
}
\end{figure}

In FIG. \ref{fig:D}, we show the individual discrimination power of ${\cal D}$ for several different observables.  In each frame only a single observable (one term) is used in the likelihood functions defining ${\cal D}$, in correspondence with TABLE \ref{tab:Data}.  

The first two frames of FIG. \ref{fig:D} show ${\cal D}$ for mass limits from colliders.
%
%
According to the Higgs masses calculated by ISAJET 7.71 the original sample contained a few models that did not respect the LEP Higgs mass limit.  These were eliminated, at varying confidence level, by the term of the likelihood function containing $m_h$.
%
%
Similarly, the second frame indicates that a few model points are excluded based on the mass limit on the lighter chargino from LEP.  When these models are paired with others the discrimination power ${\cal D}$ is very close to 1.

The two subsequent frames show ${\cal D}$ for electroweak precision variables.
%
%
Frame 3 of FIG. \ref{fig:D} reflects on the weak discrimination power of $\Delta \rho_0$: the degeneracy is broken by $\Delta \rho_0$ in only a small fraction of model pairs at a high confidence level, and the discrimination power vanishes for most of the models.  
%
%
As frame 4 of the figure shows, the anomalous magnetic moment of the muon clearly discriminates between about 100 models but its discrimination power diminishes for about half of the pairs.  This might be the result of biased selection.  The 384 MSSM model points were selected from a parameter region that can be probed by the LHC, that is from a region where superpartner masses are bounded from above.  Present collider lower limits on superpartner masses were also enforced on the sample, so superpartner masses fall into a relatively narrow window.  In this window, we expect moderate supersymmetric contributions to $a_\mu$, which accounts for a moderate discrimination power.

%
The next two frames of FIG. \ref{fig:D} show the discrimination power of two rare b-decay branching fractions $B(b \to s \gamma)$ and $B(B_s \to \mu^+\mu^-)$.  These tell a similar story to that of $\Delta a_\mu$.  Here $B(B_s \to \mu^+\mu^-)$ appears to be excellent discriminator mostly because the supersymmetric contribution to it varies sensitively with $\tan\beta$, the ratio of the two Higgs vacuum expectation values.  (A similar conclusion was reached in \cite{Isidori:2006pk}.)  This sharp variation results in a clean discrimination in model pairs where one of the points has high and the other low $\tan\beta$.

%
The discriminating power of the relic abundance of neutralinos, featured in frame 7, is also sharp.  Although $\Omega h^2$ can differentiate between only a quarter of the pairs it differentiates at high confidence levels.  This is because the theoretical prediction for $\Omega_{\tilde{Z}_1}$ tends to vary quickly (relative to the uncertainty in $\Omega$) over the parameter space.

%
The upper limit on the spin independent WIMP-proton elastic recoil cross section, shown in frame 8, is able to split a few models but its discrimination is not particularly high.  This is reflecting the well known fact that the direct detection experiments just started to become sensitive to the MSSM parameter space.  Also, the calculation of the direct detection cross section has a sizable uncertainty softening discrimination in this case.

%
The last frame of FIG. \ref{fig:D} shows gamma ray emission flux $\Phi_\gamma^{GC}$ from the galactic center as a promising discriminator.  Even with a 100 \% theoretical error, a considerable number of the model pairs are discriminated with a high confidence.  This discrimination strongly depends on the dark matter halo profile used in the calculation of the gamma ray flux (we use the popular NFW profile \cite{Navarro:1995iw}).  A conclusion quantitatively similar to ours was reached by the authors of Ref. \cite{Altunkaynak:2008ry} using the same profile.

Based on the discrimination power of ${\cal D}$ for the individual cases, we already see that the combination of the set of the nine selected experiments can result in  potentially strong discrimination.  This is indeed the case, when we calculate the combined differentiating power of all experiments using Eq.(\ref{eq:calD}).  Our final result can be simply summarized: out of the 802 model pairs 799 are distinguished at higher than 99.999 \% confidence level and only one of the model pairs cannot be resolved at better than 65 \% confidence level.  In conclusion, we can state that as far as the examined sample is concerned the inverse problem is resolved by our analysis. 

Assuming that the original set of 43026 model points was a representative sample, we expect that a more elaborate and higher statistics scan will lead to the same qualitative conclusion.  Even if, applying our approach to a higher statistics sample, one finds a higher number of degenerate model pairs our method allows for straightforward inclusion of more data.  This is important, because near future data from the Tevatron, the LHC, (in)direct detection of dark matter, b-physics and astrophysical observation will be even more constraining on supersymmetry and provide a higher level of discrimination.

\vspace{8 mm}

\section{Conclusions}

In this work we quantified the LHC inverse problem in the context of the Bayesian probability theory.  We constructed a robust discriminator, from the ratio of likelihoods of an arbitrary model pair, which powerfully distinguished essentially all examined MSSM parameter points.  To achieve this, we combined data from a range of collider, astrophysical, and low energy observables.

To reduce the inverse problem for a general theory, all available experimental information should be included.  The stochastic analysis presented above allows us to achieve this.  
In the context of our analysis the differentiating power of all, or any individual, experiments can be simply quantified in terms of confidence levels.
While we treated all experiments on the same footing, weighting can easily be introduced.

\begin{acknowledgments}

We thank the authors of Ref. \cite{Berger:2007ut} for providing us with the mass spectrum of the 283 degenerate pairs from Ref. \cite{ArkaniHamed:2005px}.  We also thank D. Carter, H. Georgi, F. Wang, and M. White for invaluable discussions on various aspects of the likelihood analysis.  This research was funded in part by the Australian Research Council under Project ID DP0877916.

\end{acknowledgments}

\bibliography{references}

\end{document}